# *In-situ* electrical and thermal transport properties of Fe$_y$Se$_{1-x}$Te$_x$ films with ionic liquid gating

Juan Xu,[1,2,†] Mingyang Qin,[2,3,†] Zefeng Lin,[2] Xu Zhang,[2] Ruozhou Zhang,[2,4] Li Xu,[2,4] Liping Zhang,[2] Qiuyan Shi,[2,4] Jie Yuan,[2,5] Beiyi Zhu,[2] Chao Dong,[6] Rui Xiong,[1] Qihong Chen,[2] Yangmu Li,[2,4,*] Jing Shi,[1,*] and Kui Jin[2,4,7]

[1]School of Physics and Technology, Wuhan University, Wuhan 430072, China
[2]Beijing National Laboratory for Condensed Matter Physics, Institute of Physics, Chinese Academy of Sciences, Beijing 100190, China
[3]Department of Materials Science and Engineering, Southern University of Science and Technology, Shenzhen 518055, China
[4]School of Physical Sciences, University of Chinese Academy of Sciences, Beijing 100049, China
[5]Key Laboratory of Vacuum Physics, School of Physical Sciences, University of Chinese Academy of Sciences, Beijing 101408, China
[6]Institute of High Energy Physics, Chinese Academy of Sciences, Beijing 100049, China
[7]Songshan Lake Materials Laboratory, Dongguan, Guangdong 523808, China

†: equal contribution
* Corresponding author: yangmuli@iphy.ac.cn, jshi@whu.edu.cn

We combine *in-situ* electrical transport and Seebeck coefficient measurements with the ionic liquid gating technique to investigate superconductivity and the normal state of Fe$_y$Se$_{1-x}$Te$_x$ (FST) films. We find that the pristine FST films feature a non-Fermi liquid temperature dependence of the Seebeck coefficient, i.e., $S/T \sim A_S \ln T$, and $A_S$ is strongly correlated with the superconducting transition temperature ($T_c$). Ionic liquid gating significantly raises $T_c$ of FST films, for which the Seebeck coefficient displays a novel scaling behavior and retains the logarithmic temperature dependence. Moreover, a quantitative relationship between the slope of $T$-linear resistivity ($A_\rho$) and $T_c$ for gated films is observed, i.e., $(A_\rho)^{1/2} \sim T_c$, consistent with previous reports on cuprates and FeSe. The scaling behaviors of $A_S$ and $A_\rho$ point to a spin-fluctuation-associated transport mechanism in gated Fe$_y$Se$_{1-x}$Te$_x$ superconductors.

Iron-based superconductors, which have attracted tremendous attention since their discovery, provide an ideal platform for studying unconventional superconductivity [1,2]. Their superconducting mechanism, however, is still under debate, e.g., ideas based on interactions among multiple orbitals fail for highly doped compounds with a single electron-type Fermi surface [3-5]; the existence of quantum criticality and its significance in superconducting pairing is elusive, hampering construction of a clear picture [6]; discrepancies between experimental details render theoretical analysis a challenging task [1]. Seeking universal behavior at the quantitative level between superconductivity and its tuning parameters in iron-based materials can lead to a promising avenue to elucidate the underlying physics.

FeSe is an archetypal iron-chalcogenide superconductor with a simple structure. Isovalent substitution of Te for Se leads to rich quantum and topological phenomena, allowing for an additional degree of freedom in controlling such as spin-orbital coupling [7-9]. More importantly, the superconducting transition temperature ($T_c$) of Fe$_y$Se$_{1-x}$Te$_x$ (FST) could be tuned over a relatively wide range through ionic liquid gating (ILG) [10], which has considerable advantages in the continuous tuning of the superconducting state by carrier doping while it has a negligible impact on crystalline structure and chemical environment [11-13]. Recently, Jiang *et al*. applied ILG to FeSe films and found a scaling between $T_c$ and the slope of the normal-state $T$-linear electrical resistivity $A_\rho$ [14]. Zhang *et al*. reported an electron effective mass associated with $T_c$ in ion-liquid-gated FeSe films [15]. These observations reveal a close connection between normal-state transport and unconventional superconductivity in iron chalcogenides.

In this work, we systematically study the electrical and thermal transport properties of the ion-liquid-gated FST films (x = 0.18, 0.47, and 0.50) by performing *in-situ* electrical resistivity ($\rho_{xx}$) and Seebeck coefficient ($S$) measurements. For the pristine FST films, $S/T$ displays a logarithmic temperature dependence, i.e., $S/T \sim A_S \ln T$, and $T_c$ monotonically correlates with the magnitude of $A_S$. For the gated FeSe films with higher $T_c$, $S/T$ collapses onto a new scaling curve. In addition, we observe $(A_\rho)^{1/2} \sim T_c$ for gated films. These observations provide evidences of spin fluctuation scattering in carrier transport for the FST system.

FST films were fabricated on SrTiO$_3$ (001) substrates by pulsed laser deposition (PLD) method [16], and crystal structures were characterized by X-ray diffraction (SmartLab X-ray diffractometer) and scanning electron microscope (Hitachi SU5000 system). To obtain the normal-state number of carriers, we patterned the film into Hall bar geometry and carried out measurements with Quantum Design Physical Property Measurement System. The *in-situ* electrical and thermal transport experiments were performed with in-house instrumentation mounted on a Montana Cryostation.

For a better gating effect, FST films were deposited on half (e.g., left side) of the substrate and Au gate electrode was patterned on the other half (e.g., right side). $C_8H_{20}NO \cdot C_2F_6NO_4S_2$, i.e., N, N-diethyl-N-methyl-N-(2-methoxyethyl) ammonium bis(trifluoromethylsulphonyl)imide, was used to bridge FST and Au electrode and supplied charged ions. A Keithley 2400 source meter was employed to apply bias voltage and measure the leakage current. The *in-situ* electrical resistivity was measured in a standard four-terminal configuration with a Keithley 2182 nanovoltmeter and a Keithley 6221 current source meter. The *in-situ* steady-state *dc* method was used for Seebeck coefficient measurement [15] with a Keithley 2182 nanovoltmeter and a Lakeshore 350 temperature controller, for which a thin-film resistance heater and a copper cold platform were utilized to achieve the temperature gradient. A complete ILG transport experiment consists of a set of gating sequences: (1) setting FST films covered by ionic liquid to a target temperature and applying a gate voltage; (2) waiting for a given period of time for injection of ions; (3) locking the ion distribution by cooling down to the ionic liquid freezing temperature about 220 K; (4) performing electrical resistivity measurement while cooling down to the base temperature; (5) performing Seebeck coefficient measurement while warming up. The continuous tuning of superconductivity for FST films was attained by adjusting the gating temperature, gating voltage, and gating time.

Figure 1(a) displays the X-ray diffraction patterns of FST films. All films are well oriented along the (00l) direction and have relatively flat surfaces, as demonstrated by the scanning electron microscopy image in Fig. 1(b). To determine the chemical concentration of the films, energy-dispersive X-ray (EDX) spectroscopy experiments are employed, as shown in Fig. 1(c). Note that EDX only provides a determination of the relative chemical ratio, and we normalize their values to the nominal $x = 0.50$ film for absolute values. The Fe concentration is found to be around 0.8 for all films, with a slightly higher value at $x = 0.18$. The corresponding $c$-axis lattice constants are 5.77, 5.91, 5.96, and 5.98 Å for $x = 0.18, 0.45, 0.47$, and 0.50, respectively, with an evolution trend consistent with earlier reports [17].

Figure 2 plots the electrical and thermal transport properties of pristine FST films. Electrical resistivity ($\rho_{xx}$) exhibits an upturn below ~ 36 K for $x = 0.18$ [18] while it stays metallic for $x = 0.45, 0.47$, and 0.50. The Seebeck coefficient, presented as $S/T$, undergoes a sign change as the temperature decreases. In Fig. 2(c), we present the Hall coefficient, $R_H = \rho_{xy}/B$, for $x = 0.45$ film. Linear Hall resistivity in magnetic fields is observed from 30 to 290 K, while a sign change of the Hall coefficient is found between 50 and 70 K, consistent with previous reports [19,20]. This sign change behavior in Seebeck and Hall coefficients was argued to originate from the multi-orbital feature of pristine FST [21,22].

The ILG process, for which charged ions are induced into FST films by gate voltage, mainly alters the number of charge carriers. Figure 3(a) illustrates the device configuration of our ILG experiments. Figure 3(b)-(d) depict the corresponding evolution of $\rho_{xx}(T)$ with gating for $x = 0.18, 0.47$, and 0.50 films, respectively. A $T$-linear resistivity behavior, which has been linked to the strange metal scattering rate [23-25], is observed for all films (~ 100 - 200 K for $x = 0.18$, and 40-100 K for $x = 0.47$ and $x = 0.50$). Because the $x = 0.18$ film displays a resistivity upturn at low temperatures, we rely on a higher temperature regime to extract its $T$-linear resistivity. Figure 3(e)-(g) compare the superconducting transition temperature $T_c$ (defined as the intersection temperature of the superconducting transition) to the slope of the $T$-linear resistivity $A_\rho$. Intriguingly, a scaling relationship $(A_\rho)^{1/2} \sim T_c$ is evident for all measured gating conditions. It is worth emphasizing that this phenomenological resistivity scaling has been discovered in La$_{2-x}$Ce$_x$CuO$_4$ films [26] and FeSe films [14] previously. This behavior provides evidence for an underlying connection between normal-state scattering and the emergence of unconventional superconductivity, whose theoretical explanations have not been fully established yet.

Figure 4 presents the Seebeck coefficient. $S/T$ undergoes a sign change for pristine films with $x = 0.18, 0.47$, and $0.50$ (dashed lines), while it becomes negative for the entire temperature range after gating (solid lines). This result is consistent with electron doping by ILG. $S/T$ of gated films features a similar temperature dependence for all gating conditions with a negative minimum value [$(S/T)_{min}$] and an $S/T \sim \ln T$ regime is discovered (~ 100 - 200 K for $x = 0.18$, and 40 -100 K for $x = 0.47$ and $x = 0.50$, consistent with the temperature range for $T$-linear resistivity). The $S/T \sim \ln T$ behavior was reported to occur in cuprate superconductors [27], iron pnictides [28], κ-HgBr [29], and heavy Fermion systems [30-32]. As demonstrated in Fig. 4(g)-(i), $S/T \sim \ln T$ can be normalized to a scaling invariant plot of $(S/T)/(S/T)_{min}$ as a function of a reduced temperature $T/T_c$.

For pristine films, the slope $A_S$ in the $S/T \sim \ln T$ regime positively correlates with the $T_c$. In Fig. 5(a), we plot a comparison between $T_c$ and $A_S$ for pristine FST films and other unconventional superconductors for which data are available (i.e., La$_{2-x}$Ce$_x$CuO$_4$ [27], Pr$_{2-x}$Ce$_x$CuO$_4$ [27], and κ-HgBr [29]). By contrast, for ion-liquid-gated FST, the correlation between $T_c$ and $A_S$ becomes inverse [Fig. 5(b)].

The $S/T \sim \ln T$ behavior in unconventional superconductors has been found in close proximity to an antiferromagnetic (AF) quantum critical point (QCP), where coupling between quasi-two-dimensional low-energy spin fluctuations and Fermi surface leads to scattering greatly enhanced "hot" regions. Since the carriers in "hot" regions dominate the transport properties, the Seebeck coefficient deviates from standard Fermi liquid behavior [33] and can be written as:

$$\frac{S}{T} = \frac{1}{e}\left(\frac{g_0^2 N'(0)}{\varepsilon_F \omega_s N(0)}\right) \ln\left(\frac{\omega_s}{\delta}\right), (1)$$

where $g_0$ is the coupling between electrons and spin fluctuations; $\omega_s$ denotes the typical energy of spin fluctuation; $N(0)$ is the density of states at the Fermi energy $\varepsilon_F$; and $\delta$ measures the deviation from the QCP in the phase space. In general, $\delta$ can be expressed as $\delta = \Gamma(p - p_c) + T$, where $p$ is an external tuning parameter of Hamiltonian (e.g., chemical concentration or ILG gate voltage); $p_c$ represents the QCP in the phase space; $\Gamma$ is an energy-scale parameter; and $T$ denotes the temperature. For relatively high temperatures, $T > \Gamma(p - p_c)$, we obtain

$$S/T \propto A_s \ln(1/T), \quad A_s = \frac{1}{e}\left(\frac{g_0^2 N'(0)}{\varepsilon_F \omega_s N(0)}\right). (2)$$

We observe this $S/T \sim \ln T$ behavior for pristine FST films (Fig. 2), indicating a significant contribution from spin fluctuations in the scattering channel. Inelastic neutron scattering for FST single crystals [7,34,35] indeed detected substantial spin fluctuations. For $x$ decreasing from 1 to 0 (i.e., FeTe to FeSe), long-range AF order disappears and polymorphic short-range spin patterns that can change as a function of temperature and Te content occur [7,34-36]. Our analysis shows that the $A_S$ for pristine FST is approximately on the same order of magnitude as those for $La_{2-x}Ce_xCuO_4$ [27], $Pr_{2-x}Ce_xCuO_4$ [27], and κ-HgBr [29], in which large spin fluctuations have been found [Fig. 5(a)].

What is surprising is the exhibition of the extraordinary scaling behavior for $S/T$ in gated FST films. $S/T$ normalized by $(S/T)_{min}$ collapses onto a single curve when plotted against $T/T_c$, which is indicative of a single mechanism dominating thermoelectric properties and the enhancement of $T_c$ although the inherent physics about this scaling behavior remain unclear. We observe that the $S/T$ of pristine FST cannot be scaled with the $S/T$ of gated FST. This discrepancy likely originates from a carrier contribution change, for which a sign change of $S/T$ is seen for pristine FST but not for gated FST films with higher $T_c$. The change in carrier types is consistent with the previous reports that high-$T_c$ iron chalcogenide superconductors have only electron pockets [3,37].

$A_S$ has an inverse correlation with $T_c$ for the gated samples, as shown in Fig. 5(b), in contrast to that for the pristine films [Fig. 5(a)]. What causes the difference? From Eq. (3), spin fluctuation energy $\omega_s$ is proportional to $W$, the bandwidth of the conduction electrons assuming to be an approximate constant. $A_S$ is relevant to both $g_0$ and energy-band parameters $\varepsilon_F$, $N(0)$, and $N'(0)$, which are influenced by carrier concentration ($n$). For the pristine FST films, $n$ is in the same order of magnitude for the amount of isovalent Te-substitution of Se [19], similar to the situation of $La_{2-x}Ce_xCuO_4$ [26] and $Pr_{2-x}Ce_xCuO_4$ [38] in the heavily overdoped regime. Under such circumstances, $A_S$ is mainly dominated by $g_0$ and it is positively correlated with $T_c$, as shown in Fig. 5(a). However, in gated FST samples, the variation of $n$ induced by gating cannot be neglected [15]. Here, we tentatively estimate the gating effect on the energy-band parameters near Fermi surface based on the multiorbital band structure of FST. The electron contribution, mainly near the M point, can be approximated by elliptical Fermi surfaces [39-41]. In the two-dimensional approximation, both of $N(0)$ and $N'(0)$ are independent of $n$, $\varepsilon_F \propto n$, so $A_S$ is a function of $n^{-1}$. In the three-dimensional limit, $N(0) \propto n^{1/3}$, $N'(0) \propto n^{-1/3}$, and $\varepsilon_F \propto n^{2/3}$, and $A_S$ is thus proportional to $n^{-2/3}$. In both cases, $A_S$ is inversely correlated with $n$, consistent with the observation in Fig. 5(b), since higher $T_c$ corresponds to an increase of $n$ (see Fig. 4(a)-(c) and related analyses). The $A_S$ in gated films is mainly dominated by the change of band parameters.

The scaling between strange metal scattering and superconducting transition temperature has been linked to a prominent presence of spin fluctuations in cuprate superconductors and iron chalcogenides [14,26]. $T$-linear resistivity in our gated samples is in line with this argument. We find that $(A_\rho)^{1/2} \sim T_c$ in gated FST is similar to that of the gated FeSe [14], and their overall slope is consistent with those reported for $La_{2-x}Ce_xCuO_4$, $La_{2-x}Sr_xCuO_4$, and $(TMTSF)_2PF_6$ [26]. This universal behavior, again, points to the important role of spin fluctuations in the scattering channel.

In summary, we combine the ionic liquid gating with *in-situ* transport measurements to systematically study the FST films. We observe non-Fermi liquid Seebeck coefficient $S/T \sim A_S \ln T$ and $T$-linear resistivity $\rho_{xx} \sim A_\rho T$. We obtain $(A_\rho)^{1/2} \sim T_c$ and a scaling relationship between $A_S$ and $T_c$. These scaling behaviors provide evidence for the effect of spin fluctuations, and shed new light on the mechanism of unconventional superconductivity.

Acknowledgement:
This work was supported by the National Key Basic Research Program of China (Grants No. 2021YFA0718700, 2018YFB0704102, 2017YFA0303003, 2017YFA0302902, 2022YFA1603903 and 2021YFA0718802), the National Natural Science Foundation of China (Grants No. 11927808, 11834016, 118115301, 119611410, 11961141008, 61727805, 11961141002 and 12274439), the Key Research Program of Frontier Sciences, Chinese Academy of Sciences (CAS) (Grants QYZDB-SSW-SLH008 and QYZDY-SSW-SLH001), CAS Interdisciplinary Innovation Team, the Strategic Priority Research Program (B) of CAS (Grants No. XDB25000000 and XDB33000000), CAS through the Youth Innovation Promotion Association (Grant No. 2022YSBR-048), the Beijing Natural Science Foundation (Grant No. Z190008) and the Key-Area Research and Development Program of Guangdong Province (Grant No. 2020B0101340002). M.Q. thanks for the support from the China Postdoctoral Science Foundation (Grant No. 2022M711497).

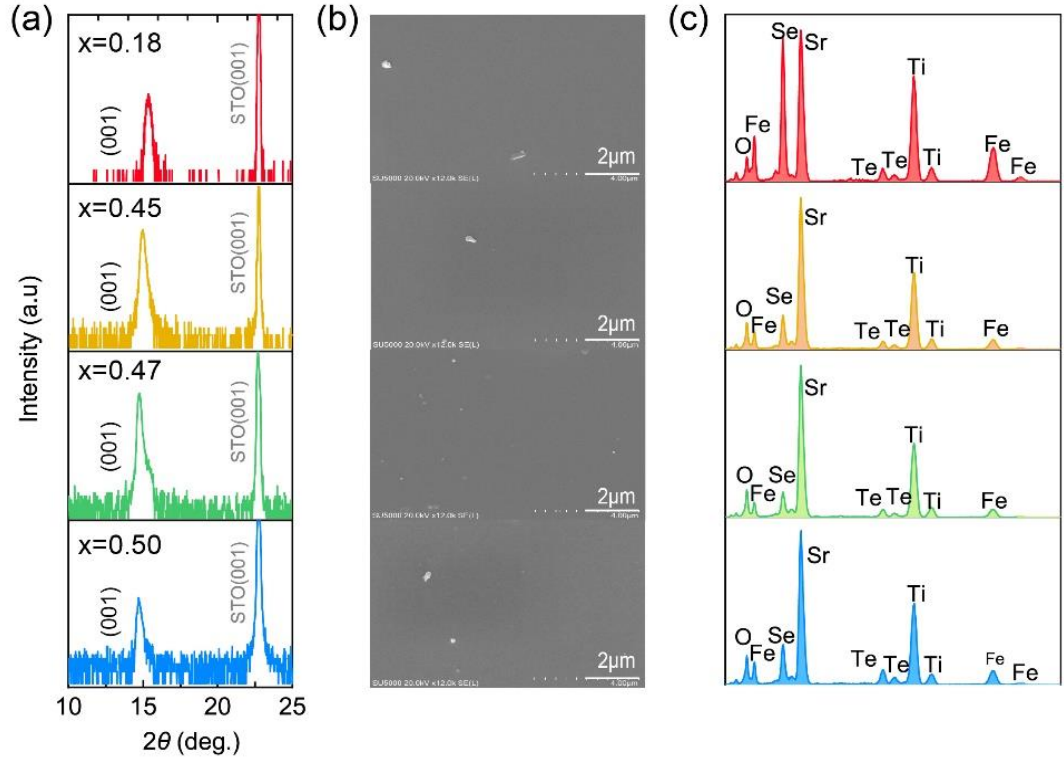

FIG.1. Structural characterization and chemical composition for $Fe_ySe_{1-x}Te_x$. (a) Out-of-plane XRD Bragg peaks for films with $x$ = 0.18, 0.45, 0.47, and 0.50 films and STO substrates. (b) SEM tomographs showing micrometer-scale flat surfaces. (c) EDX spectra for chemical compositions.

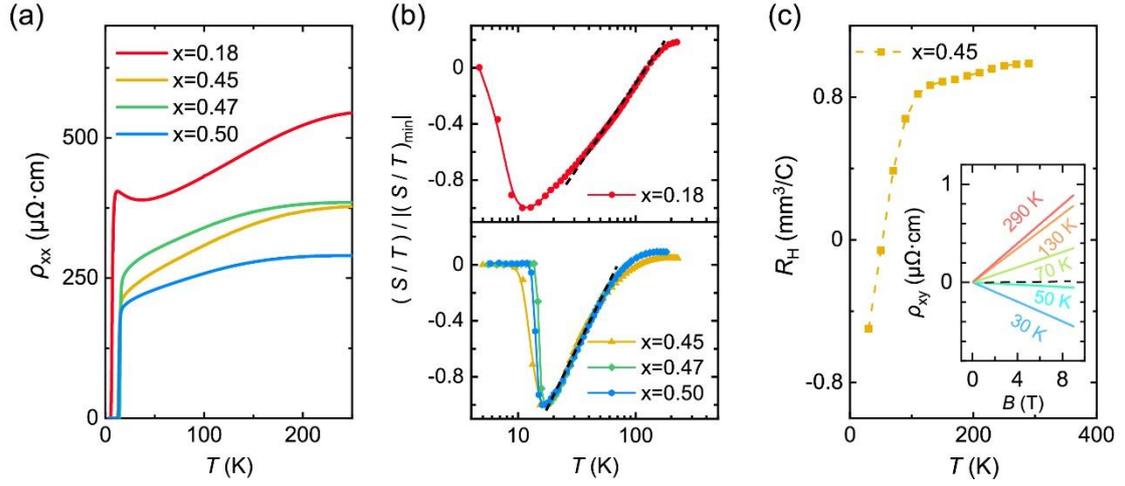

FIG.2. Electrical and thermal transport properties of pristine $Fe_ySe_{1-x}Te_x$ films. (a) Temperature dependence of electrical resistivity. (b) Normalized Seebeck coefficient (S), plotted as $S/T$ versus temperature on a semi-log scale. The dashed lines depict the $S/T \sim \ln T$ region. (c) Hall coefficient as a function of temperature for $x = 0.45$ film, where the inset shows a linear Hall resistivity in magnetic fields.

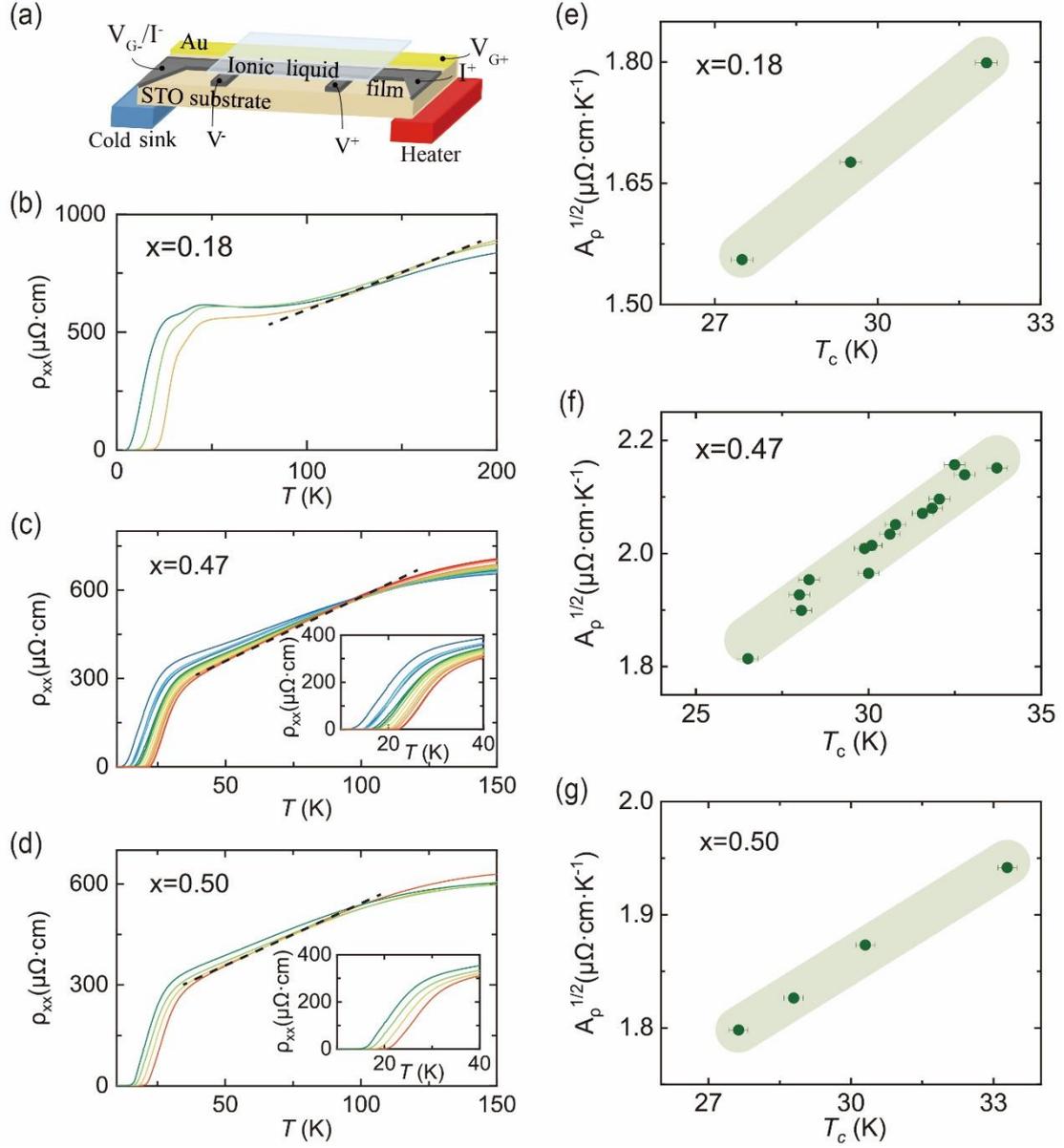

FIG.3. Electrical transport properties of Fe$_y$Se$_{1-x}$Te$_x$ films with ionic liquid gating (ILG). (a) ILG device configuration. (b)-(d) Electrical resistivity as a function of temperature for with $x$ = 0.18, 0.47, 0.50 with various gating conditions. An enhancement of $T_c$ is seen in all films. The dashed lines indicate $T$-linear resistivity. The insets in (c) and (d) present zoomed-in plots of resistivity near the superconducting transition. (e)-(g) Comparison between the slope of the $T$-linear electrical resistivity $A_\rho$ and $T_c$ for all films.

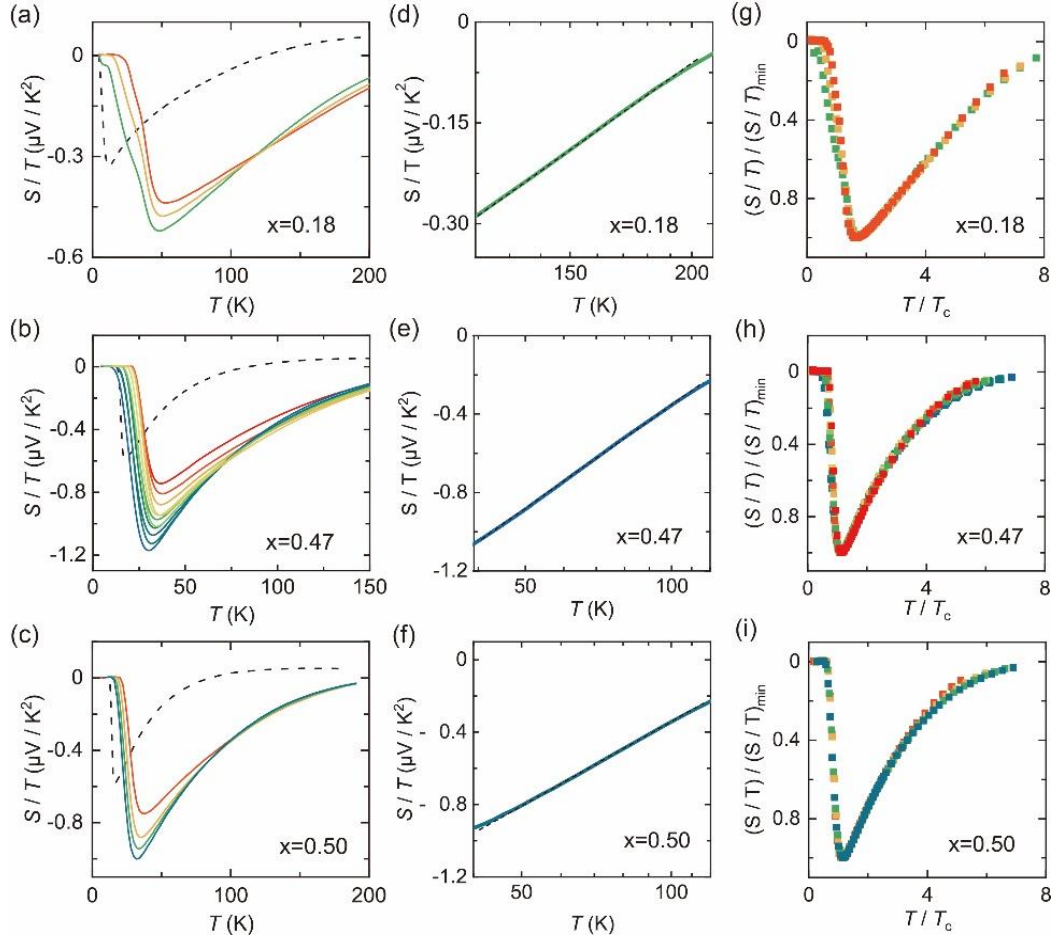

FIG.4. Thermal transport properties of $Fe_ySe_{1-x}Te_x$ films with ionic liquid gating (ILG). (a)-(c) Temperature dependence of Seebeck coefficient, plotted as $S/T$ for $x = 0.18, 0.47, 0.50$ films. The dashed lines show the temperature dependence of $S/T$ for pristine films. (d)-(f) Representative $S/T \sim \ln T$ regime for gated films in (a)-(c). (g)-(i) Normalized $S/T$ data as a function of $T/T_c$.

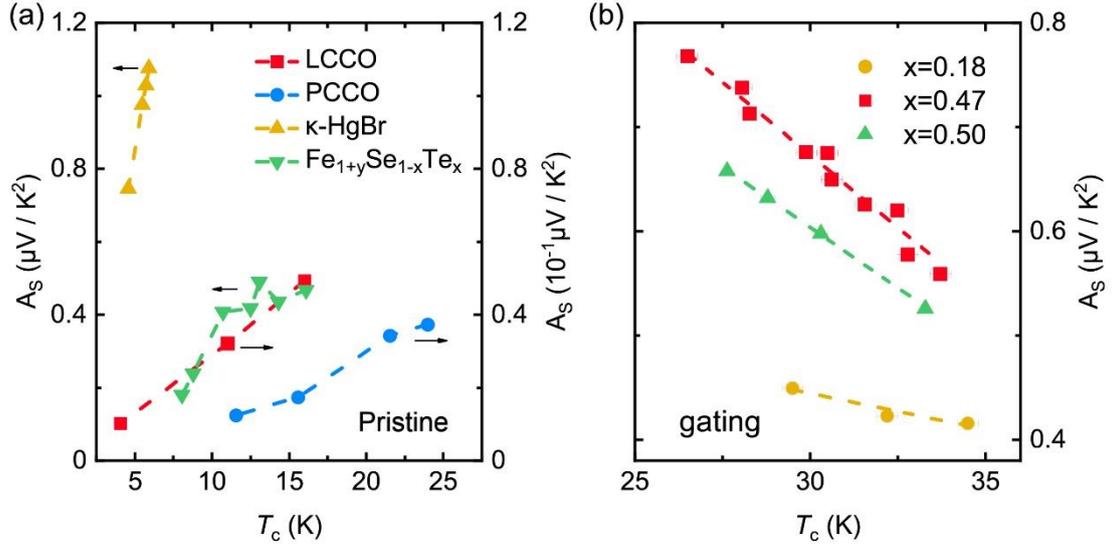

FIG.5. Comparison between slope $A_S$ of the $S/T \sim \ln T$ regime and $T_c$ for a variety of superconductors. (a) Comparison for pristine films. Data of pristine $Fe_ySe_{1-x}Te_x$ films are taken from Fig. 4 and our additional measurements. Data of chemically substituted $La_{2-x}Ce_xCuO_4$ and $Pr_{2-x}Ce_xCuO_4$, and data of hydrostatic pressurized κ-HgBr are taken from refs. [27,29]. (b) Comparison for ion-liquid-gated $Fe_ySe_{1-x}Te_x$ films. Data were taken from Fig. 4.